\documentclass{article}

\usepackage[final]{neurips_2020_ml4ps}

\usepackage[utf8]{inputenc} % allow utf-8 input
\usepackage[T1]{fontenc}    % use 8-bit T1 fonts
\usepackage{hyperref}       % hyperlinks
\usepackage{url}            % simple URL typesetting
\usepackage{booktabs}       % professional-quality tables
\usepackage{amsfonts}       % blackboard math symbols
\usepackage{amsmath}
\usepackage{microtype}      % microtypography
\usepackage{color}
\usepackage{graphicx}
\usepackage{graphbox}
\usepackage{aas_macros}

\title{Probabilistic Mapping of Dark Matter \\ by Neural Score Matching}

\author{%
  Benjamin Remy\\
  AIM, CEA, CNRS \\ 
  Universit\'e Paris-Saclay \\ 
  Universit\'e Paris Diderot \\
  Sorbonne Paris Cit\'e \\ 
  F-91191 Gif-sur-Yvette, France \\
  \texttt{benjamin.remy@cea.fr} \\
  \And
  Fran\c{c}ois Lanusse \\
  AIM, CEA, CNRS \\ 
  Universit\'e Paris-Saclay \\ 
  Universit\'e Paris Diderot \\
  Sorbonne Paris Cit\'e \\ 
  F-91191 Gif-sur-Yvette, France \\
  \texttt{francois.lanusse@cea.fr} \\
  \And
  Zaccharie Ramzi \\
  CEA (Neurospin and Cosmostat) \\ Inria (Parietal)\\
  Gif-sur-Yvette, France \\
  \texttt{zaccharie.ramzi@inria.fr}
  \And
  Jia Liu \\
  Berkeley Center for Cosmological Physics \\
  University of California \\
  Berkeley, CA 94720, USA \\
  \texttt{jialiu@berkeley.edu}\\
  \And
  Niall Jeffrey$^{a,b}$ \\
  $^{a}$Laboratoire de Physique de\\ l'Ecole Normale Sup\'erieure, ENS \\
  Universit\'e PSL, CNRS\\
  Sorbonne Universit\'e\\ Universit\'e de Paris, 
Paris, France\\
  $^{b}$University College London\\
  Gower St, London, UK \\
  \texttt{niall.jeffrey@phys.ens.fr}\\
  \And
  Jean-Luc Starck\\
  AIM, CEA, CNRS \\ 
  Universit\'e Paris-Saclay \\ 
  Universit\'e Paris Diderot \\
  Sorbonne Paris Cit\'e \\ 
  F-91191 Gif-sur-Yvette, France \\
  \texttt{jstarck@cea.fr}
}

\begin{document}

\maketitle

\begin{abstract}
The Dark Matter present in the Large-Scale Structure of the Universe is invisible,
but its presence can be inferred through the small gravitational lensing effect it 
has on the images of far away galaxies. By measuring this lensing effect on a large number of galaxies it is possible to reconstruct maps of the Dark Matter distribution on the sky. This, however,
represents an extremely challenging inverse problem due to missing data and noise 
dominated measurements. In this work, we present a novel methodology for addressing
such inverse problems by combining elements of Bayesian statistics, analytic 
physical theory, and a recent class of Deep Generative Models based on Neural Score Matching.
This approach allows to do the following: (1) make full use of analytic cosmological theory
to constrain the 2pt statistics of the solution, (2) learn from cosmological simulations any 
differences between this analytic prior and full simulations, and (3) obtain samples from the 
full Bayesian posterior of the problem for robust Uncertainty Quantification. We present 
an application of this methodology on the first deep-learning-assisted Dark Matter map 
reconstruction of the Hubble Space Telescope COSMOS field.
\end{abstract}

\section{Introduction}

Deep Learning methods have proven extremely efficient at solving a wide range of inverse problems such as
deconvolution, denoising, super-resolution, inpainting, etc. As a concrete example, deep learning models are
now state-of-the-art in MRI \citep{Ramzi2020BenchmarkingDatasets}. Yet, to apply these methods in the Physical Sciences, two particular
preoccupations always remain:  How do we robustly quantify uncertainties on a deep learning solution? How do we combine deep learning with existing and trusted physical models? In this work, we present a methodology allowing us to address both of these questions in a consistent Bayesian framework and apply it to a challenging instance of inverse problems from the field of Cosmology: mapping the distribution of Dark Matter in the Large-Scale Structure of the Universe from its weak gravitational lensing effect.  Several factors make this problem a perfect example:
\begin{itemize}
    \item[-] The mapping problem is ill-posed, there is \textit{no unique solution} and it cannot be solved without making some assumptions on the solution: we want a method that doesn't return a single solution, but a range of solutions covering all possible Dark Matter maps compatible with the observations and our a priori knowledge.
    \item[-] Analytic cosmological \textit{theory can predict some properties of Dark Matter maps}, in particular their two-point functions: we want to use as much as possible analytic modeling to constrain these solutions, i.e. we do not want to use deep learning in the regimes where theory applies.
    \item[-] Numerical cosmological \textit{simulations can sample realistic realisations of Dark Matter maps}, in all of their complexity:  we want to restrict the use of deep learning to only modeling \textit{residuals} between analytic theory and full simulations.
\end{itemize}

\section{Mapping the Invisible Dark Matter through its Gravitational Effects}

% Things we want to stay in this section.
%  - Weak lensing principles
%  - Rich amount of information in the matter distribution, point for instance to HOS papers
%  - What makes the problem difficult in practice, present it as an inverse problem
%  - Metion theory and simulations.
%  - Introduce state-of-the-art / existing techniques, and highlight the gap between uncertainties and deep learning
The field of cosmology is investing multiple-billion dollars in space- and ground-based telescopes, with the aim to unveil the fundamental nature of our Universe: dark energy, Dark Matter, cosmic inflation, and Einstein's theory of gravity. The matter distribution---where and how all the matter is positioned throughout space---holds the key to these puzzles. The main difficulty stems from the fact that visible matter, the stars and galaxies, only accounts for 15\% of total matter, and the remaining 85\% is in the form of invisible Dark Matter. 
%In the past decade, the observational technique of weak gravitational lensing has advanced rapidly, allowing us to begin mapping out the dark universe using distant galaxies. 
Under Einstein's theory of gravity, the path of a light ray is bent when it passes by massive objects, an effect similar to that of magnifying glasses (and hence the name ``gravitational lensing''). Using distant galaxies as our background image, by measuring the distortion of the galaxy shape, % or magnification of the galaxy brightness, 
we can infer the total matter distribution.

In practice, our task is to infer the matter distribution map, known as a convergence $\boldsymbol{\kappa}$ map or simply as a mass map, from measurements of individual galaxy ellipticities $\boldsymbol{\epsilon}$. These ellipticities are affected by gravitational lensing and can be used as estimators of the lensing signal. From a Bayesian perspective, the shape-to-mass $\boldsymbol{\epsilon}\to\boldsymbol{\kappa}$ inference problem can be quantified as, 
\begin{equation}
    p(\boldsymbol{\kappa} | \boldsymbol{\epsilon}, \mathcal{M}) \propto p(\boldsymbol{\epsilon} |  \boldsymbol{\kappa}, \mathcal{M}) \ p(\boldsymbol{\kappa} | \mathcal{M}), \ \  \label{eq:bayes}
\end{equation}
where $\mathcal{M}$ is the cosmological model that encapsulates all the fundamental physics of interest. The likelihood factor $p(\boldsymbol{\epsilon} |  \boldsymbol{\kappa}, \mathcal{M})$, which describes how light-rays are bent by gravity, how measurements are affected by noise, and accounts for missing observational data. All these factors contribute to making the problem non-invertible, but are all well-understood. The prior distribution however, $p(\boldsymbol{\kappa} | \mathcal{M})$, which describes the detailed matter distribution under any model $\mathcal{M}$, lacks a full analytic expression. Cosmological theory can provide an analytic Gaussian prior which is accurate but only on large scales.

While a number of approximate methods  have been adopted to quantify this prior term on smaller scales~\cite{marshall_mass_maps, glimpse2016, alsing2016hierarchical, alsing2017cosmological, jeffrey2018, price_maps}, at present the only accurate physical model of convergence maps for Dark Matter mapping rely on large-scale numerical simulations (presented in \cite{deeplearning_shirasaki, deepmass}), such as the \texttt{MassiveNuS} simulations used in this work~\cite{Liu2018MassiveNuS}. As described in \cite{deepmass}, forward simulation lets us draw samples from $p(\boldsymbol{\kappa} | \mathcal{M})$. Our goal in this work is to sample the full posterior distribution $p(\boldsymbol{\kappa} | \boldsymbol{\epsilon}, \mathcal{M})$.

%Some recent methods based on deep learning \citep{Shirasaki2018,deepmass} have been able to take advantage of these forward simulations, by using them to build datasets on which a deep neural network is trained to directly solve the mass-mapping problem. The main drawback of these approaches however is that the neural network output does not allow for robust Uncertainty Quantification, i.e. accessing the full posterior distribution $p(\boldsymbol{\kappa} | \boldsymbol{\epsilon}, \mathcal{M})$.

\section{Neural Score Matching for Bayesian Inverse Problems}

%In many applications, the likelihood term in the Bayesian formulation of the inverse problem is fairly well known. This is for instance the case for our gravitational lensing problem in  \autoref{eq:bayes}, but is also true for a wide range of applications from MRI
%to radio-interferometry. The more problematic term is the prior $p(x)$ which is typically a lot more
%complicated to evaluate. % Up until a few years ago, practical solutions for solving these inverse
%problems required simple but tractable signal priors, one of the most consequential of them being the sparsity prior.

%However, 
With the fast development of Deep Generative Models, it has recently become possible to learn high dimensional probability distributions from data, opening the prospect of data-driven deep priors. In particular, generative models with explicit likelihoods, like PixelCNNs \citep{pixelcnn++} and Normalizing Flow \citep{Rezende2015VariationalFlows}, can directly learn, evaluate, and sample any $p(\boldsymbol{x})$ given enough training examples. However, a significant realization is that in the perspective of using such generative models as part of a Bayesian inverse problem such as \autoref{eq:bayes}, i.e. alongside a likelihood term, there is no need for a full normalized probability distribution function $p(\boldsymbol{x})$. Instead, all we need to sample the posterior with modern gradient-based inference techniques such as Hamiltonian Monte-Carlo  \cite[HMC,][]{betancourt2018conceptual} or Black-Box Variational Inference \cite{JMLR:v14:hoffman13a}, is to have access to the \textit{score} $\nabla_{\boldsymbol{x}} \log p(\boldsymbol{x})$. Very recent work \citep{Song2019,Song2020} has demonstrated that generative models based on learning this score can easily reach GAN quality samples and are both more scalable and easier to train.% (in this work, we wok on 320x320 images). %Hence the idea developed in this work, to learn the score of the prior term and use it as part of a gradient-based sampling scheme to retrieve posterior samples.

% in the perspective of using these generative models for sampling from a Bayesian Inverse Problem posterior, one realisation is that it is not necessary to have access to the full normalized probability distribution function, only the score $\nabla_x \log p(x)$ is required in modern gradient-based inference techniques such as Hamiltonian Monte-Carlo \cite{betancourt2018conceptual} or Black-Box Variational Inference \cite{JMLR:v14:hoffman13a}. Hence the idea developed in this work, to learn the score of the prior term and use it as part of a gradient-based sampling scheme to retrieve posterior samples.

% \FL{Mention that this works in much higher dimensions than otherwise.}
% \FL{Mention that this works in much higher dimensions than otherwise.}

\paragraph{Denoising Score Matching} As originally identified in \cite{Vincent2011} and \cite{Alain2013}, the score of a given target distribution $P$ can be modeled using a Denoising Auto-Encoder (DAE), i.e. by introducing an auto-encoding function $\boldsymbol{r}_\theta: \mathbb{R}^{n} \times \mathbb{R} \mapsto \mathbb{R}^{n}$ trained to reconstruct a true $\boldsymbol{x} \sim P$ given a noisy version $\boldsymbol{x}' = \boldsymbol{x} + \boldsymbol{n}$ with $\boldsymbol{n} \sim \mathcal{N}(0, \sigma^2 \boldsymbol{I})$ under an $\ell_2$ loss. An optimal denoiser $\boldsymbol{r}^\star$ would then be achieved for:
\begin{equation}
    \boldsymbol{r}^\star(\boldsymbol{x}, \sigma) = \boldsymbol{x} + \sigma^2 \nabla_{\boldsymbol{x}} \log p_{\sigma^2}(\boldsymbol{x})
\end{equation}
where $p_{\sigma^2} = p \ast \mathcal{N}(0, \sigma^2)$. In other words, the optimal denoiser is closely related to the score we wish to learn and when the noise variance $\sigma^2$ tends to zero, should exactly match the score of the target density.  In practice to learn this score efficiently, we adopt the noise-conditional Denoising Score Matching (DSM) technique proposed by \citep{Lim2020}, and train a model to minimize the loss:
\begin{equation}
    \mathcal{L}_{DSM} = \underset{\begin{subarray}{c}
  \boldsymbol{u} \sim \mathcal{N}(0, I) \\
  \sigma \sim \mathcal{N}(0, s^2)
  \end{subarray}}{\mathbb{E}} \parallel \boldsymbol{u} + \sigma \boldsymbol{r}_{\theta}(\boldsymbol{x} + \sigma \boldsymbol{u}, \sigma) \parallel_2^2 \label{eq:dsn}
\end{equation}
In this expression, note that $\boldsymbol{r}_{\theta}(\boldsymbol{x} + \sigma \boldsymbol{u} , \sigma)$ directly models the score $\nabla_{\boldsymbol{x}} \log p_{\sigma^2}$ and is conditioned on the level of noise $\sigma$ used to corrupt the input. In practice, the network is a 3-scale U-net with residual blocks composed of convolutions followed by a batch normalisation.

% An important point to stress for the following, is that the optimum for this loss is achieved by:
% \begin{equation}
%     \boldsymbol{r}^*(\boldsymbol{x}, \sigma) = \nabla_{\boldsymbol{x}} \log p_{\sigma^2}(\boldsymbol{x})
% \end{equation}
% i.e. the model is learning the score of the target distribution convolved with a Gaussian of 
% scale $\sigma$. 
% At this point, we should have explained how we can learn a score from a distribution

\paragraph{Annealed Hamiltonian-Monte Carlo Sampling} Given the noise-conditional neural scores learned with the procedure described above, it is now possible to use a variety of inference methods to access the Bayesian posterior. In this work, we adopt an annealed HMC procedure which provides an efficient way to obtain independent samples from the target posterior despite the high dimensionality of the problem. 
We consider a tempered version of a target posterior density:
\begin{equation}
    \log p_{\sigma^2}(\boldsymbol{x} | \boldsymbol{y}) = \log p_{\sigma^2}(\boldsymbol{y} |\boldsymbol{x}) + \log p_{\sigma^2} (\boldsymbol{x})
\end{equation}
where $\sigma^2$ plays the role of the inverse temperature found in thermodynamic annealing methods, and we recognise the $\log p_{\sigma^2} (\boldsymbol{x})$ term modeled by the score network $\boldsymbol{r}_\theta(., \sigma)$. At high temperatures, i.e. large $\sigma$, the tempered distribution tends to a wide Gaussian, all modes merge together, and an HMC with a diagonal mass matrix can efficiently navigate the posterior distribution. This distribution is then gradually annealed to low temperatures and parallel independent chains progressively moves towards different points in the target distribution. Once we have reached 
low temperature, we retrieve the last sample from each parallel chain, yielding a batch of independent samples of the posterior distribution. This procedure is similar to the Annealed Langevin diffusion proposed in \cite{Song2019}.

% Sampling in high-dimensions, even with Hamiltonian Monte-Carlo can be extremely challenging 
% when the mass matrix of the problem is unknown, and not diagonal. Several methods have been proposed to speed up sampling speed in HMC in these situations, in particular Riemannian Manifold Hamiltonian Monte-Carlo and its deep-learning powered derivation NeuTra \cite{hoffman2019neutralizing}. In this work, instead of trying to achieve an efficient sampling over the target manifold of a given
% chain, we adopt the opposite approach of annealing sampling of a large number of chains in parallel. 
% \FL{Explain this better}close to that proposed in \cite{Song2019}.

% We consider a tempered version of our target density:
% \begin{equation}
%     \log p_{\sigma^2}(\boldsymbol{x} | \boldsymbol{y}) = \log p_{\sigma^2}(\boldsymbol{y} |\boldsymbol{x}) + \log p_{\sigma^2} (\boldsymbol{x})
% \end{equation}
% where $\sigma^2$ plays the role of the inverse temperature found in classical annealing \FL{add reference and 1/kT}.

% \FL{Mention the calibration part}
% What do we want to say
%   - The annealing part, we need to explain that HMC sampling when we don't have an appropriate mass matrix is extremely slow, so batch annealing allows us to rapidly sample multiple realisation
%   - The calibration part, despite the fact that we don't have access to the target density
%   - The annealing schedule 
% \FL{this is in contrast to \cite{Song2019} who rely on an uncalibrated Annealed Langevin sampling}

\begin{figure}
    \centering
    \includegraphics[width=\textwidth]{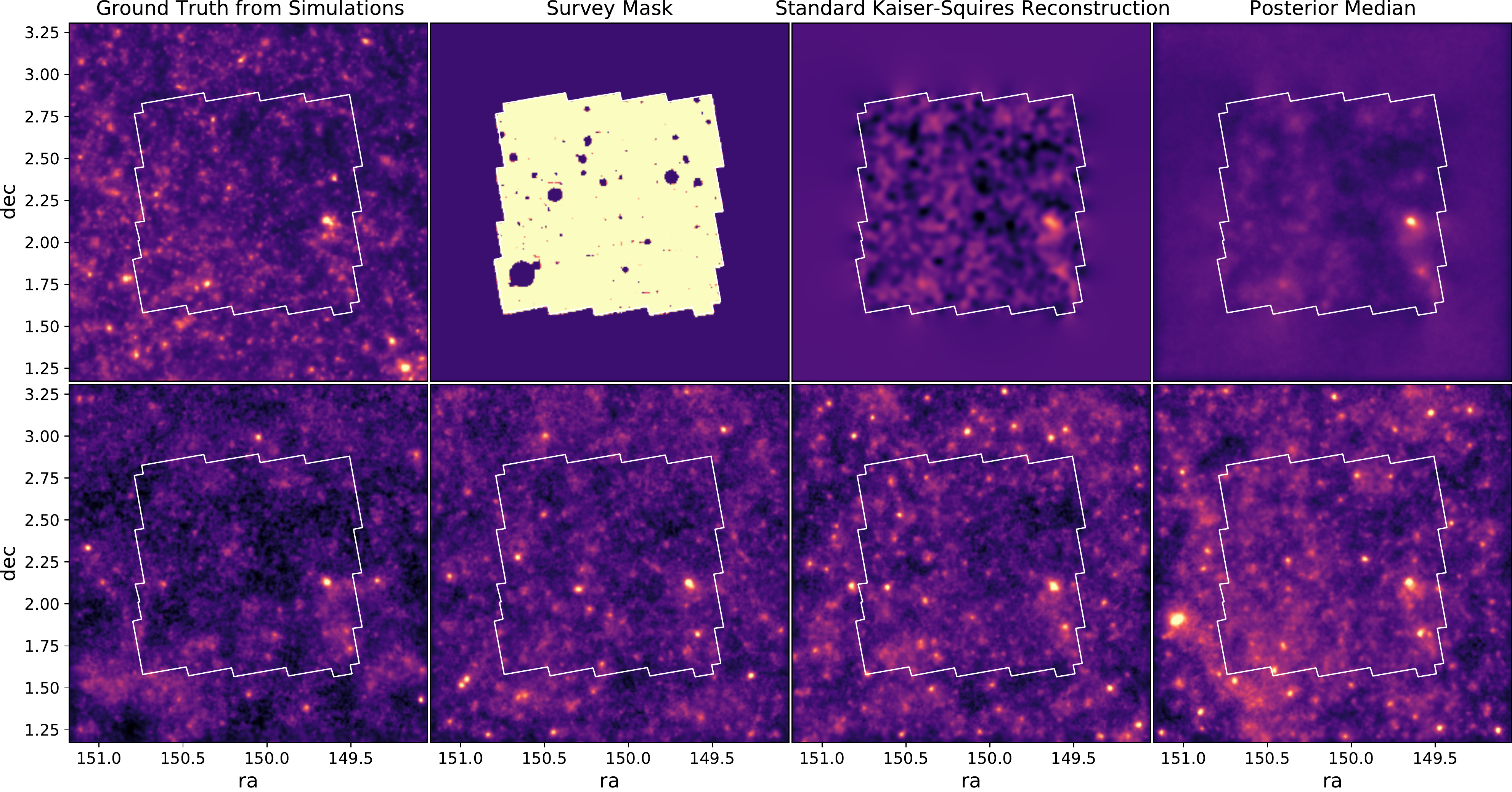}
    \caption{\small Reconstruction of Dark Matter maps on simulated Hubble Space Telescope lensing measurements. The top row
    shows from left to right, true map from the \texttt{MassiveNuS} simulations, the mask showing the extent of the COSMOS survey, a  reconstructed map using a conventional Kaiser-Squires method, and the median of our posterior samples. The bottom row
    shows multiple samples from the posterior obtained with our procedure. }
    \label{fig1}
    %\vspace{-0.5cm}
\end{figure}

\section{Hybrid Analytic and Neural Score Modeling for Mass-Mapping}

The method presented in the previous section was based on learning the full score from simulations. However, some properties of the Dark Matter maps are known from analytic cosmology models. We now present how we can directly combine analytic modeling and deep Neural Score Matching. \\
The two-point statistics of weak lensing fields can be accurately predicted from theory over a range of scales, 
and this forms the basis of most lensing-based cosmological constraints to date. Only on small scales do the 
theoretical models start to deviate from reality. Since these two-point properties are so well-known, we do not
wish to leave it up to the neural network to learn them, and instead want to directly impose this prior knowledge on our 
solution.
This prior on $\boldsymbol{\kappa}$ can be expressed as a Gaussian with a covariance $\mathbf{S}$ which becomes diagonal in Fourier space:
\begin{equation}
    p_{th}(\boldsymbol{\kappa}) = \frac{1}{ \sqrt{ \det 2 \pi \mathbf{S}}} \exp \left( -\frac{1}{2} \boldsymbol{\kappa}^\dagger \mathbf{S}^{-1} \boldsymbol{\kappa} \right)
\end{equation}
Using this Gaussian prior alone in a reconstruction would yield Gaussian constrained realisations \citep{Hoffman91} or the conventional Wiener filter if the maximum a posteriori solution is the target \cite{Lahav94}. But we know the convergence field is \textit{not} Gaussian, so we aim to learn the difference between the true distribution of convergence maps and a Gaussian model. In our Score Matching framework, we propose the following decomposition of the score of the full prior $p(\boldsymbol{\kappa})$:
\begin{equation}
    \nabla_{\boldsymbol{\kappa}} \log p(\boldsymbol{\kappa}) = \nabla_{\boldsymbol{\kappa}} \log p_{th}(\boldsymbol{\kappa}) + \boldsymbol{r}_\theta(\boldsymbol{\kappa}, \nabla_{\boldsymbol{\kappa}} \log p_{th}(\boldsymbol{\kappa}))
\end{equation}
where $\boldsymbol{r}_\theta$ is a score network, tasked with modeling the difference between true and Gaussian scores. This network is trained by Denoising Score Matching, simply by adapting \autoref{eq:dsn} to add the theory score to the network output.
% Just like for 
% conventional Denoising Score Matching, we can train the model on simulations by adapting \autoref{eq:dsn} to Residual Denoising Score Matching:
% \begin{equation}
%     \mathcal{L}_{RDSM} = \underset{\begin{subarray}{c}
%   \boldsymbol{u} \sim \mathcal{N}(0, I) \\
%   \sigma \sim \mathcal{N}(0, s^2)
%   \end{subarray}}{\mathbb{E}} \parallel \boldsymbol{u} + \sigma *\left[ \boldsymbol{r}(\boldsymbol{x} + \sigma*\boldsymbol{u}, \sigma, \nabla \log p_{th}(\boldsymbol{x}+ \sigma*\boldsymbol{u}))  + \nabla \log p_{th}(\boldsymbol{x}+ \sigma*\boldsymbol{u}) \right] \parallel_2^2 \label{eq:residual_dsn}
% \end{equation}
The main advantage of this approach is that we decrease our reliance on the neural score model to a minimum, lowering the demand on the model complexity.% of the model.% In particular, we wouldn't need to increase the depth of the model if we increased the size of the fields being reconstructed as these additional large scales can be accurately handled by the Gaussian prior.

\section{Results}

\begin{figure}
    \centering
    \includegraphics[width=0.77\textwidth, align=c]{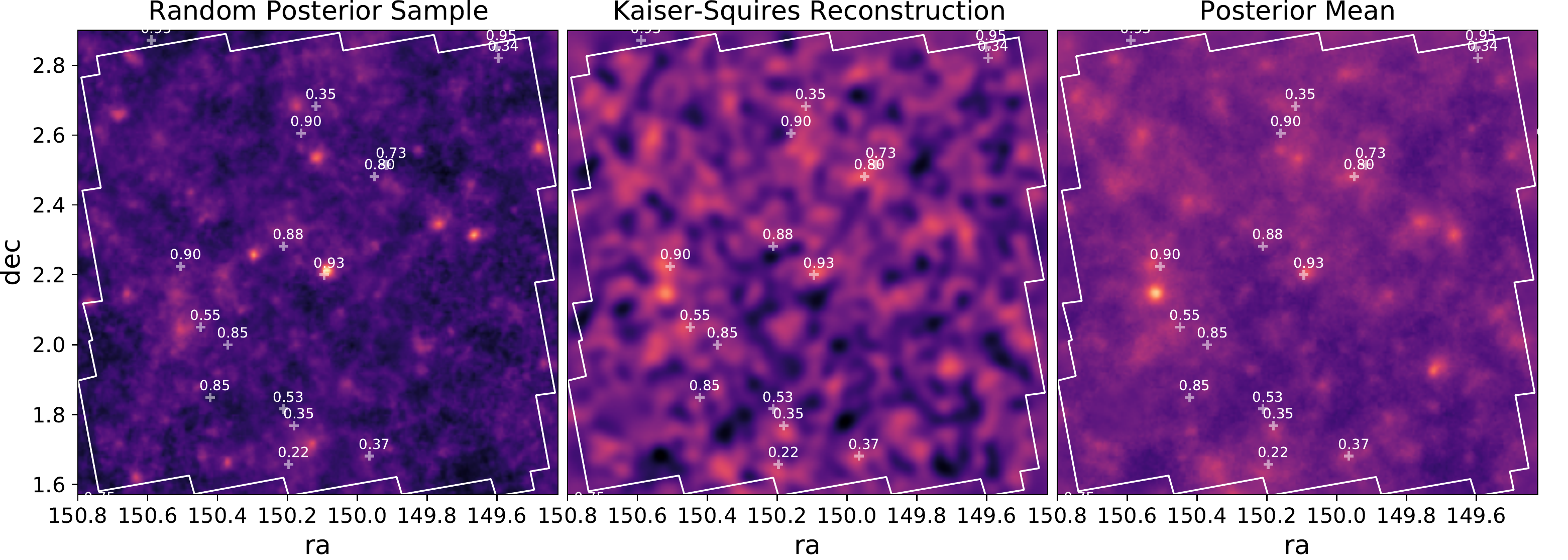}~
    \includegraphics[width=0.22\textwidth, align=c]{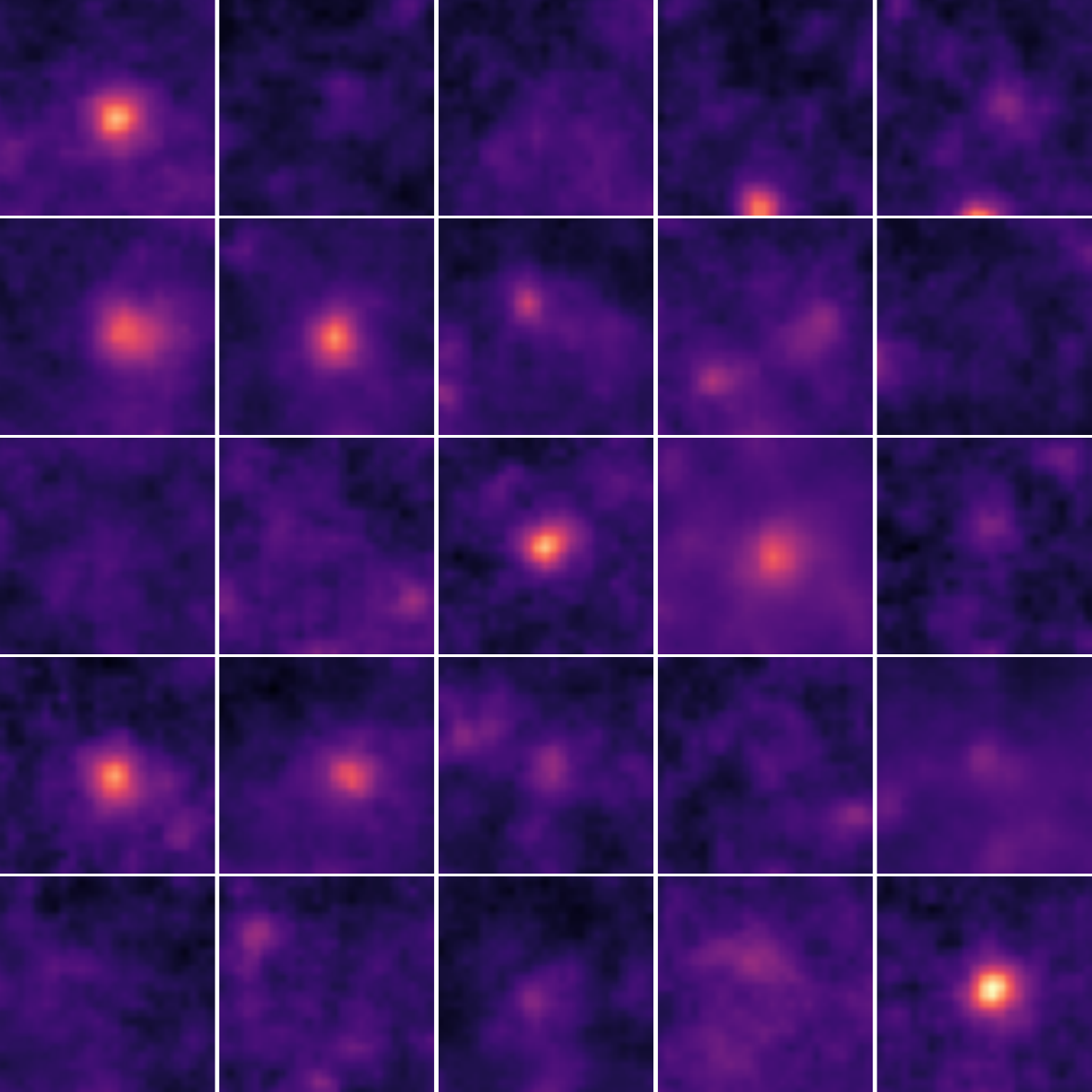}
    \caption{\small Dark Matter map reconstruction of HST COSMOS survey. We provide both a random sample of the posterior and the posterior mean, to be compared to a conventional Kaiser-Squires reconstruction. Positions of known X-ray clusters and their redshifts are indicated in white. The rightmost plot shows cutouts from independent posterior samples of the central region of the map where a massive cluster is known to be present. As can be seen, the posterior is bi-modal in this region, with the cluster present only in some of the samples.}
    \label{fig:cosmos_data}
    %\vspace{-0.5cm}
\end{figure}

To demonstrate our methodology, we present the first deep-learning assisted, probabilistic, Dark Matter map reconstruction of the Hubble Space Telescope (HST) Cosmic Evolution Survey (COSMOS) field \cite{Scoville_2007}, using data described in \cite{Schrabback_2010}, and basing our prior on the \texttt{MassiveNuS} suite of simulations \citep{Liu2018MassiveNuS}. 
%In our analysis, we use simulated mass maps from the \texttt{MassiveNuS} fiducial run which assume a flat-$\Lambda \rm{CDM}$ universe. 
The simulations contain $1024^3$~particles in a  (512~$h^{-1}$Mpc)$^3$ box with periodic boundary conditions. % It tracks the the matter distribution from early universe $z=100$ to today, accurately captures the nonlinear evolution.
10,000 mass maps are generated by ray-tracing through the simulation boxes, following the physics of gravitational lensing. We make use of the maps for source galaxies at $z=1$, which corresponds to the mean redshift of our COSMOS lensing catalog. %The maps are originally 512$^2$ pixels and 3.5$^2$ deg$^2$ in size, but randomly cropped to 320$^2$ pixels regions for augmentation.
%. During our training, we use 90\% of the available maps, randomly select a 320$^2$ pixels region of each map, and further augment the maps by randomly flipping and rotating them. 
Along with these maps, we use the corresponding theoretical power spectrum to build the analytic prior $p_{th}(\boldsymbol{\kappa})$ for training the score network.
\paragraph{Validation on simulation} We first emulate from \texttt{MassiveNuS} simulations a mock COSMOS lensing catalog. We use the actual distribution of galaxies in that survey 
to create a binary mask representing where lensing is actually measured, and we 
%Masked regions typically correspond to the presence of bright stars, or other artefacts that prevent the reliable measurement of galaxy shapes in parts of the survey.
mock observations by adding realistic levels of noise matching the COSMOS bright galaxy sample of \citep{Schrabback_2010}.
%to the \texttt{MassiveNuS} shear maps, using a standard deviation $\sigma_e=0.27$ by galaxy ellipticity component, and an averaged effective galaxy number density of 18.75 gal./arcmin$^2$, which corresponds to the COSMOS bright galaxy sample of \citep{Schrabback_2010}.
\autoref{fig1} illustrates the results of our method on this simulated dataset. 
%The top row shows from left to right: the true convergence map from simulations we aim to recover, the survey mask showing where we actually have access to lensing data in bright color, what the conventional Kaiser-Squires \citep{ks93} recovers, and finally the median of our posterior, computed over 128 samples. 
The bottom row shows 4 independent samples from the posterior. Note that because the data is very noisy, independent posterior samples can exhibit significant variability, except where a strong signal is present. This is for instance the case of the massive cluster visible in the bottom right corner of the field, visible even in the map recovered by a traditional Kaiser-Squires \citep{ks93} reconstruction. This structure being highly significant, it therefore appears virtually in all posterior samples. Also note how realistic each posterior sample looks compared to the simulation, demonstrating the quality of our score matching model on these 320x320 pixel images.
%\FL{We also want to say that the statistics of the real sims are well captured}

\paragraph{COSMOS field reconstruction} Finally, we present the application of our method to the real COSMOS field, based on the bright galaxy sample described in \citep{Schrabback_2010}. Not only does our posterior mean map represents the highest quality map ever made of the COSMOS field, we have access to the full posterior for
interpreting these results. As an example, \autoref{fig:cosmos_data} on the right shows cutouts from independent posterior samples of the central region of the map where a massive cluster is known to be present. We see that the posterior is bi-modal in this region, with the cluster present in some maps and not in others. Measuring the frequency of the cluster appearing in posterior samples would yield a robust quantification of the significance of this structure.

These results demonstrate the benefits of our proposed methodology, which optimally makes use of our physical models and knowledge (through the known likelihood term, theoretical prior on large scales, numerical simulations for small scale non-gaussian prior) and allows us to access a full posterior distribution in high dimension.
% \section{Conclusion}

% We have presented a new approach to solve mass mapping inverse problem, allowing for robust uncertainty quantification, by accessing the full posterior distribution. 

\section*{Broader Impact}
The methodology described here finds many useful applications outside of cosmology. Of particular interest, the same method can be applied to medical imaging (e.g. MRI) and help physicians assess the significance of particular parts of an image before making a diagnostic. We believe this work does not entail any negative consequences or ethical issues.

%\section*{References}
%\bibliographystyle{unsrt}
\bibliographystyle{apalike}

\bibliography{des_lfi}

\begin{thebibliography}{}

\bibitem[Alain and Bengio, 2013]{Alain2013}
Alain, G. and Bengio, Y. (2013).
\newblock {What regularized auto-encoders learn from the data generating
  distribution}.
\newblock {\em 1st International Conference on Learning Representations, ICLR
  2013 - Conference Track Proceedings}, 15:3743--3773.

\bibitem[Alsing et~al., 2017]{alsing2017cosmological}
Alsing, J., Heavens, A., and Jaffe, A.~H. (2017).
\newblock Cosmological parameters, shear maps and power spectra from cfhtlens
  using bayesian hierarchical inference.
\newblock {\em Monthly Notices of the Royal Astronomical Society},
  466(3):3272--3292.

\bibitem[Alsing et~al., 2016]{alsing2016hierarchical}
Alsing, J., Heavens, A., Jaffe, A.~H., Kiessling, A., Wandelt, B., and
  Hoffmann, T. (2016).
\newblock Hierarchical cosmic shear power spectrum inference.
\newblock {\em Monthly Notices of the Royal Astronomical Society},
  455(4):4452--4466.

\bibitem[Betancourt, 2018]{betancourt2018conceptual}
Betancourt, M. (2018).
\newblock A conceptual introduction to hamiltonian monte carlo.

\bibitem[Hoffman et~al., 2013]{JMLR:v14:hoffman13a}
Hoffman, M.~D., Blei, D.~M., Wang, C., and Paisley, J. (2013).
\newblock Stochastic variational inference.
\newblock {\em Journal of Machine Learning Research}, 14(4):1303--1347.

\bibitem[{Hoffman} and {Ribak}, 1991]{Hoffman91}
{Hoffman}, Y. and {Ribak}, E. (1991).
\newblock {Constrained Realizations of Gaussian Fields: A Simple Algorithm}.
\newblock {\em apjl}, 380:L5.

\bibitem[{Jeffrey} et~al., 2018]{jeffrey2018}
{Jeffrey}, N., {Abdalla}, F.~B., {Lahav}, O., {Lanusse}, F., {Starck}, J.~L.,
  {Leonard}, A., {Kirk}, D., {Chang}, C., {Baxter}, E., and {Kacprzak}, T.
  (2018).
\newblock {Improving weak lensing mass map reconstructions using Gaussian and
  sparsity priors: application to DES SV}.
\newblock {\em \mnras}, 479(3):2871--2888.

\bibitem[{Jeffrey} et~al., 2020]{deepmass}
{Jeffrey}, N., {Lanusse}, F., {Lahav}, O., and {Starck}, J.-L. (2020).
\newblock {Deep learning dark matter map reconstructions from DES SV weak
  lensing data}.
\newblock {\em \mnras}, 492(4):5023--5029.

\bibitem[{Kaiser} and {Squires}, 1993]{ks93}
{Kaiser}, N. and {Squires}, G. (1993).
\newblock {Mapping the Dark Matter with Weak Gravitational Lensing}.
\newblock {\em \apj}, 404:441.

\bibitem[{Lahav} et~al., 1994]{Lahav94}
{Lahav}, O., {Fisher}, K.~B., {Hoffman}, Y., {Scharf}, C.~A., and {Zaroubi}, S.
  (1994).
\newblock {Wiener Reconstruction of All-Sky Galaxy Surveys in Spherical
  Harmonics}.
\newblock {\em apjl}, 423:L93.

\bibitem[{Lanusse} et~al., 2016]{glimpse2016}
{Lanusse}, F., {Starck}, J.-L., {Leonard}, A., and {Pires}, S. (2016).
\newblock {High resolution weak lensing mass mapping combining shear and
  flexion}.
\newblock {\em \aap}, 591,A2.

\bibitem[Lim et~al., 2020]{Lim2020}
Lim, J.~H., Courville, A., Pal, C., and Huang, C.-W. (2020).
\newblock {AR-DAE: Towards Unbiased Neural Entropy Gradient Estimation}.

\bibitem[Liu et~al., 2018]{Liu2018MassiveNuS}
Liu, J., Bird, S., Matilla, J. M.~Z., Hill, J.~C., Haiman, Z., Madhavacheril,
  M.~S., Spergel, D.~N., and Petri, A. (2018).
\newblock {MassiveNuS: Cosmological massive neutrino simulations}.
\newblock {\em Journal of Cosmology and Astroparticle Physics}, 2018(3).

\bibitem[{Marshall} et~al., 2002]{marshall_mass_maps}
{Marshall}, P.~J., {Hobson}, M.~P., {Gull}, S.~F., and {Bridle}, S.~L. (2002).
\newblock {Maximum-entropy weak lens reconstruction: improved methods and
  application to data}.
\newblock {\em \mnras}, 335(4):1037--1048.

\bibitem[Price et~al., 2019]{price_maps}
Price, M.~A., Cai, X., McEwen, J.~D., Pereyra, M., and Kitching, T.~D. (2019).
\newblock Sparse bayesian mass-mapping with uncertainties: local credible
  intervals.
\newblock {\em Mon. Not. Roy. Astron. Soc.}, 492(1):394--404.

\bibitem[Ramzi et~al., 2020]{Ramzi2020BenchmarkingDatasets}
Ramzi, Z., Ciuciu, P., and Starck, J.~L. (2020).
\newblock {Benchmarking MRI reconstruction neural networks on large public
  datasets}.
\newblock {\em Applied Sciences (Switzerland)}, 10(5).

\bibitem[Rezende and Mohamed, 2015]{Rezende2015VariationalFlows}
Rezende, D.~J. and Mohamed, S. (2015).
\newblock {Variational inference with normalizing flows}.
\newblock {\em 32nd International Conference on Machine Learning, ICML 2015},
  2:1530--1538.

\bibitem[Salimans et~al., 2017]{pixelcnn++}
Salimans, T., Karpathy, A., Chen, X., and Kingma, D.~P. (2017).
\newblock Pixelcnn++: Improving the pixelcnn with discretized logistic mixture
  likelihood and other modifications.
\newblock {\em CoRR}, abs/1701.05517.

\bibitem[Schrabback et~al., 2010]{Schrabback_2010}
Schrabback, T., Hartlap, J., Joachimi, B., Kilbinger, M., Simon, P., Benabed,
  K., Bradač, M., Eifler, T., Erben, T., Fassnacht, C.~D., and et~al. (2010).
\newblock Evidence of the accelerated expansion of the universe from weak
  lensing tomography with cosmos.
\newblock {\em Astronomy and Astrophysics}, 516:A63.

\bibitem[Scoville et~al., 2007]{Scoville_2007}
Scoville, N., Abraham, R.~G., Aussel, H., Barnes, J.~E., Benson, A., Blain,
  A.~W., Calzetti, D., Comastri, A., Capak, P., Carilli, C., and et~al. (2007).
\newblock Cosmos: Hubble space telescope observations.
\newblock {\em The Astrophysical Journal Supplement Series}, 172(1):38–45.

\bibitem[{Shirasaki} et~al., 2019]{deeplearning_shirasaki}
{Shirasaki}, M., {Yoshida}, N., and {Ikeda}, S. (2019).
\newblock {Denoising weak lensing mass maps with deep learning}.
\newblock {\em \prd}, 100(4):043527.

\bibitem[Song and Ermon, 2019]{Song2019}
Song, Y. and Ermon, S. (2019).
\newblock {Generative Modeling by Estimating Gradients of the Data
  Distribution}.
\newblock (NeurIPS).

\bibitem[Song and Ermon, 2020]{Song2020}
Song, Y. and Ermon, S. (2020).
\newblock Improved techniques for training score-based generative models.

\bibitem[Vincent, 2011]{Vincent2011}
Vincent, P. (2011).
\newblock {A connection between scorematching and denoising autoencoders}.
\newblock {\em Neural Computation}, 23(7):1661--1674.

\end{thebibliography}

\end{document}